\documentclass[twoside,11pt]{article}

\usepackage{blindtext}

\usepackage{dmlr2e}
\usepackage{amsmath}
\usepackage{bm}
\usepackage{appendix}
\usepackage{cleveref}
\usepackage{xcolor}

\crefname{section}{§}{§§}

\usepackage{lastpage}
\dmlrheading{23}{2024}{1-\pageref{LastPage}}{01/24; Revised 03/24}{05/24}{21-0000}{Pandya and Yang} 
\def\openreview{\url{https://openreview.net/forum?id=BTiIPYEXWY}} 

\ShortHeadings{Learning Galaxy Intrinsic Alignment Correlations}{Pandya and Yang}
\firstpageno{1}

\begin{document}

\title{Learning Galaxy Intrinsic Alignment Correlations}


\author{\name Sneh Pandya \email pandya.sne@northeastern.edu \\
       \addr Department of Physics, Northeastern University, Boston, MA 02115, USA\\
       NSF AI Institute for Artificial Intelligence and Fundamental Interactions (IAIFI)
       \AND
       \name Yuanyuan Yang \email yang.yuanyu@northeastern.edu \\
       \addr Khoury College of Computer Sciences, Northeastern University, Boston, MA 02115, USA
       \AND
       \name Nicholas Van Alfen \email vanalfen.n@northeastern.edu \\
       \addr Department of Physics, Northeastern University, Boston, MA 02115, USA
       \AND
       \name Jonathan Blazek \email j.blazek@northeastern.edu \\
       \addr Department of Physics, Northeastern University, Boston, MA 02115, USA
       \AND
       \name Robin Walters \email r.walters@northeastern.edu \\
       \addr Khoury College of Computer Sciences, Northeastern University, Boston, MA 02115, USA
       }


\maketitle

\begin{abstract}
The intrinsic alignments (IA) of galaxies, regarded as a contaminant in weak lensing analyses, represents the correlation of galaxy shapes due to gravitational tidal interactions and galaxy formation processes. As such, understanding IA is paramount for accurate cosmological inferences from weak lensing surveys; however, one limitation to our understanding and mitigation of IA is expensive simulation-based modeling. In this work, we present a deep learning approach to emulate galaxy position-position ($\xi$), position-orientation ($\omega$), and orientation-orientation ($\eta$) correlation function measurements and uncertainties from halo occupation distribution-based mock galaxy catalogs. We find strong Pearson correlation values with the model across all three correlation functions and further predict aleatoric uncertainties through a mean-variance estimation training procedure. $\xi(r)$ predictions are generally accurate to $\leq10\%$. Our model also successfully captures the underlying signal of the noisier correlations $\omega(r)$ and $\eta(r)$, although with a lower average accuracy. We find that the model performance is inhibited by the stochasticity of the data, and will benefit from correlations averaged over multiple data realizations. Our code will be made open source upon journal publication.
\end{abstract}


\section{Introduction}
\label{sec:intro}

Intrinsic Alignment (IA) describes the correlation between galaxy shapes themselves, as well as the correlation between galaxy shapes and the distribution of underlying dark matter,  a hypothetical form of matter that does not emit, absorb, or reflect light, yet it is thought to constitute approximately 85\% of the matter in the universe and accordingly influences the structure and behavior of galaxies and galaxy clusters. These correlations pose a significant challenge in cosmological analyses. 
While IA offers insights into the large-scale structure of the universe, it is also a contaminant for weak gravitational lensing signals (see \citealt{TROXEL20151} for a detailed review). Weak lensing, an effect where light is deflected by gravitational fields, serves as a critical tool for studying matter distributions and cosmological constraints. However, its subtlety makes it susceptible to contamination by IA, potentially leading to significant systematic errors in signal interpretation (e.g.\ \citealt{hirata04}, \citealt{10.1093/mnras/stv2615}, \citealt{PhysRevD.100.103506}).
IA analyses, traditionally modeled by analytic approaches which fail to capture alignments in the full non-linear regime, have recently turned to simulation models for more accurate descriptions. These methods, however, suffer from computational expense and would benefit from efficient modeling on GPUs.



Machine learning (ML) techniques, especially neural networks (NNs), have found wide success in the sciences with the advent of high performance computing and large datasets, particularly in astrophysics and cosmology \citep{dvorkin2022machine}. 
Of particular interest is the potential for NNs to emulate expensive N-body \citep{Jamieson_2023} and magneto-hydrodynamic simulations \citep{Rosofsky_2023}.

In this project, we present a novel deep learning method to model both IA correlation amplitudes and uncertainties for galaxy IA statistics. Our proposed solution is a NN-based encoder-decoder architecture, trained on a wide array of galaxy catalogs derived from N-body simulations and augmented with Halo Occupation Distribution (HOD) techniques. For a given set of HOD parameters, the model is capable of simultaneous inference for all three correlation functions constructed from galaxy positions and orientations. This emulator is intended to expedite and streamline the modeling process, thereby enabling comprehensive data analysis and efficient Monte Carlo based parameter inference.

\subsection{Related Work}

Several previous works have constructed simulation-based emulators for cosmological statistics, with a focus on matter and/or galaxy density.
\cite{Zhai_2019} constructed Gaussian process based emulators based on the AEMULUS Project's N-body simulations for nonlinear galaxy clustering.
\cite{Kwan_2023} similarly used a Gaussian process based emulator, HOD modeling, and the Mira-Titan Suite of N-body simulations to predict galaxy correlation functions, building on earlier work from the same group \citep{2010ApJ...713.1322L}.
The BACCO simulation project (\citealt{2021arXiv210414568A}, \citealt{2021MNRAS.506.4070A}) built NN emulators to include nonlinear and baryonic effects from simulations.
These projects emulate various cosmological statistics from simulations, but do not include IA. \cite{Jagvaral_2022} and \cite{2023arXiv231211707J} developed generative models trained on the IllustrisTNG-100 simulation \citep{nelson2021illustristng} to emulate IA in hydrodynamic simulations, but these models do not emulate statistics.
This work is the first attempt at emulating galaxy-IA correlation statistics using simulated galaxy catalogs.

\section{Data \& Method}

\begin{figure}
\centering
\includegraphics[width=\textwidth]{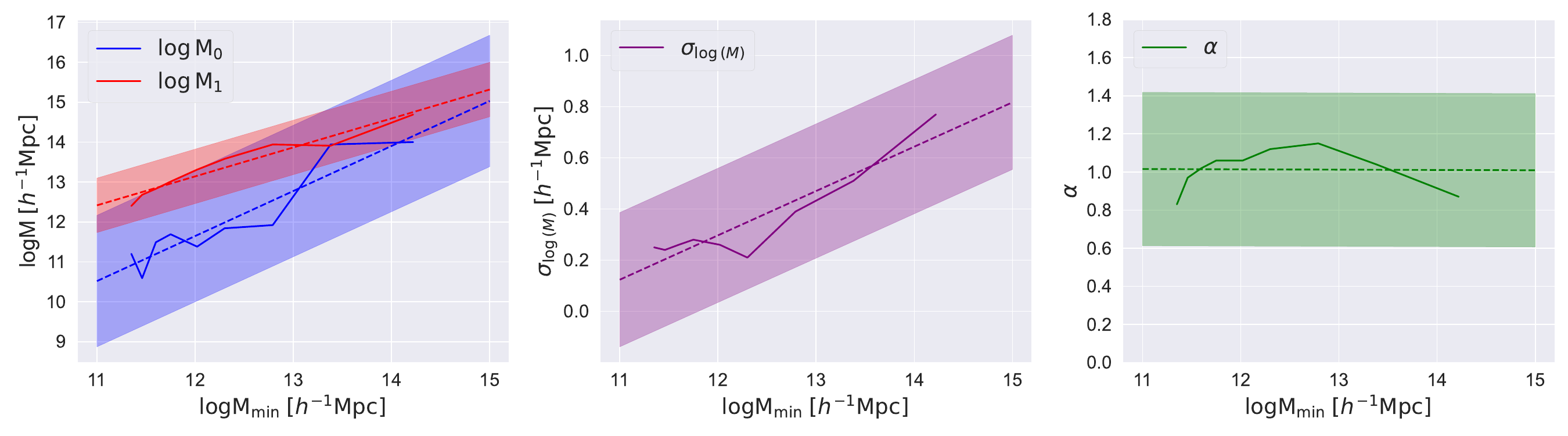}
\caption{We generate uniform random values for the four occupation parameters, excluding $\log \mathrm{M_{min}}$. These values are based on a linear relationship with $\log \mathrm{M_{min}}$, serving as a central line. The range for random values extends $4\times\mathrm{RMSE}$ surrounding this line. To clarify the visualization, $\sigma_{\log(\mathrm{M})}$ is displayed separately from other mass variables. Each panel presents published data from \citet{Zheng_2007} as a solid line, while the dotted line illustrates the linear fit to $\log \mathrm{M_{min}}$, with the shaded area indicating the range for uniform random value selection for each parameter. Not shown here are the two IA parameters, $\mu_{\rm cen}$ and $\mu_{\rm sat}$, which both vary uniformly on the range $[-1,1]$ with no relation to these five occupation parameters.}
\label{fig:parameter-draw}
\end{figure}

\subsection{Simulated Dataset}
\label{sec:sim}

For training the emulator, we must generate extensive galaxy catalogs that reflect realistic universes, further extracting galaxy correlation statistics from them. These, in combination with HOD model input parameters, formulate the dataset used for supervised learning.

We generate catalogs of galaxies using the {\tt halotools} Python package. This package was created to produce mock galaxy catalogs using a HOD model-based method by populating existing catalogs of dark matter halos \citep{Hearin_2017}. 
This model is extended in \citealt{vanalfen_2023} to include a component for aligning galaxies to model IA; it is this version of {\tt halotools} that we use to generate the training data for the emulator.

The HOD models encompass two populations of galaxies: central galaxies which are the most massive galaxies at the center of the larger parent halo (an isolated dark matter halo), and satellite galaxies which are less massive galaxies within that parent halo positioned elsewhere. Subhalos are defined as a dark matter halo existing within another dark matter halo.
An HOD model is built using an occupation component which determines the number of galaxies to be populated within a given dark matter halo, and a phase space component which determines the galaxy positions and velocities. The extension to IA adds a galaxy alignment component, the strength of which can be set to depend on other galaxy properties.

For the purposes of this emulator, 
we use the {\tt halotools} built-in occupation models \texttt{Zheng07Cens} and \texttt{Zheng07Sats} for central and satellite galaxies, respectively. These occupation model components populate dark matter halos following equations 2, 3, and 5 of \citet{Zheng_2007}. 
Central galaxies are placed at the center of the parent halo, and for simplicity, we use a subhalo phase space model for satellite galaxies which places satellites at the position of their respective subhalos. We adopt a central alignment model aligning central galaxies with their parent dark matter halo's major axis, and for satellites, a radial alignment model, aligning them along the vector from their central galaxy. Both alignment models are stochastic.


Correlation functions are measured for the position--position ($\xi(r)$), position--orientation ($\omega(r)$), and orientation--orientation ($\eta(r)$) correlations in 20 bins between galaxies in the simulation at a minimum separation of $0.1\, h^{-1}{\rm Mpc}$ up to a maximum separation of $16\, h^{-1}{\rm Mpc}$. 
The correlation function estimators are mathematically defined as
\begin{equation}
    \xi(r) = \frac{DD(r)}{RR(r)} - 1, \qquad \omega(r) = \langle|\hat{e}(\boldsymbol{x}) \cdot \hat{r}|^2 \rangle - \frac{1}{3}, \qquad \eta(r) = \langle|\hat{e}(\boldsymbol{x}) \cdot \hat{e}(\boldsymbol{x} + \boldsymbol{r})|^2\rangle - \frac{1}{3}
\end{equation}
where $\boldsymbol{x}$ is the position vector of a galaxy, $DD(r)$ denotes the number of galaxy pairs separated by $r$, $RR(r)$ is the expected number of pairs for a random sample, and $\hat{e}(\boldsymbol{x})$ is a 3D orientation unit vector of a galaxy. 
In future work, we plan to extend the maximum range of this correlation, but for purposes of building this model we chose this maximum separation 
as correlations at this distance can still be measured quickly. In general, galaxies at $r \leq 1\, h^{-1} {\rm Mpc}$ are in the ``1-halo regime''(galaxies within the same halo) and galaxies outside this range are in the ``2-halo regime'' (galaxies residing in separate halos).

The input model parameters to build the HOD model and generate the datasets are the central alignment strength, $\mu_{\rm cen}$, the satellite alignment strength, $\mu_{\rm sat}$, and five occupation components: $\log{M_{min}}$, $\sigma_{\log{M}}$, $\log{M_0}$, $\log{M_1}$, and $\alpha$ all described in \citet{Zheng_2007}. To get full coverage, we first attempted a Latin Hypercube on these parameters following \citealt{Kwan_2023}, using the range of published values for the \texttt{Zheng07Cens} and \texttt{Zheng07Sats} occupation models in {\tt halotools} \citep{Zheng_2007}.
While this provides an efficient way to generate unique combinations of parameters,
we find that it can create mock galaxy catalogs with unphysical characteristics, such as $\xi(r)$ correlation amplitudes with much higher values than is typical. Specifically, the galaxy--galaxy $\xi(r)$ correlation functions were often more than $100$ times greater than the dark matter $\xi(r)$ correlation function, suggesting unrealistic universes.
Although the individual values chosen for each parameter fall within range of something physical, it is clear that the Latin Hypercube method can sample regions of parameter space that are not of interest.
To avoid distorting the model with unphysical training  data, we restrict ourselves to regions of parameter space produce realistic galaxy populations.


In selecting parameter values, we use the nine points from Table 1 in \citet{Zheng_2007} to establish a linear relationship between $\log{M_{min}}$ and each parameter, calculating the RMSE for these fits. We then generate a sequence of evenly spaced $\log{M_{min}}$ values within the SDSS's range for each parameter. For each $\log{M_{min}}$ value, we randomly select values for the other parameters from a range centered on the linear fit, spanning $4*{\rm RMSE}$ as shown in Figure \ref{fig:parameter-draw}. $\mu_{\rm cen}$ and $\mu_{\rm sat}$ are sampled uniformly from the interval $[-1, 1]$. 
We populate a catalog for each set of parameters and obtain the correlations to compose the dataset of $116383$ samples, further splitting it into a $70\%$ train, $10\%$ validation, and $20\%$ test sets. The training data was generated using a combination of 2.4 GHz Intel E5-2680 CPUs and 2.1 GHz Intel Xeon Platinum 8176 CPUs. The simulations were parallelized across 150 cores, split evenly to allow simultaneous calculation of the correlation functions.

\subsection{Model Architecture}
\label{sec:arch}

\begin{figure}
    \centering
    \includegraphics[width=\textwidth]{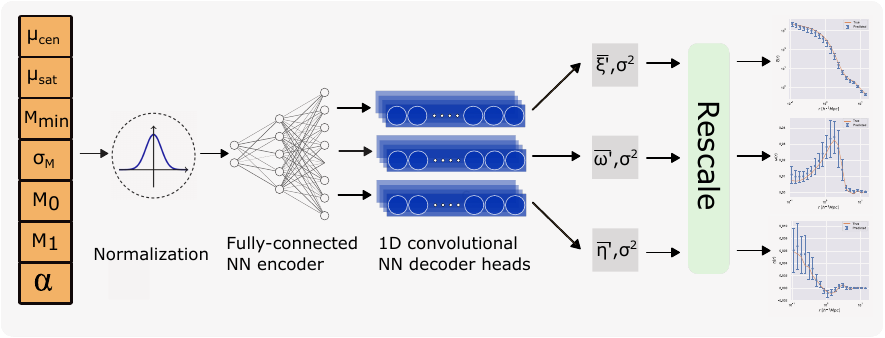}
    \caption{Model Pipeline. A logarithm factor is implied for all masses $M$. The HOD input model parameters are normalized before entering the 7-layer deep fully-connected encoder. The encoder expands the dimensionality of the input until the decoder stage, which features four 1D convolutional layers which learn the individual local correlations present in the output correlation functions, $\xi'$, $\omega'$, and $\eta'$. These are then re-scaled back to their original values. A detailed description of the encoder-decoder architecture is shown in Appendix \ref{sec:AppendixB}.}
    \label{fig:pipeline}
\end{figure}

Our objective is to construct an NN that embodies the HOD simulation and maps a 7-dimensional input vector of cosmological parameters outlined in \cref{sec:sim} to the correlation functions $\xi(r)$, $\omega(r)$, and $\eta(r)$, each comprising 20 bins. We further seek to predict the aleatoric uncertainties on the correlation amplitudes to capture the stochastic nature of HOD modeling through a mean-variance estimation training procedure \citep{MVE}. This is important for $\omega(r)$ and particularly $\eta(r)$, which are inherently very noisy statistics due to the significant effects of galaxy orientation noise in correlations \citep{Bernstein_2002}. Mathematically, this mapping can be represented by a function $f: \mathbb{R}^7 \rightarrow \mathbb{R}^{40} \times \mathbb{R}^{40} \times \mathbb{R}^{40}$. We utilize \texttt{PyTorch} to construct an architecture that encompasses a fully-connected NN shared encoder head and three 1D convolutional NN decoder heads that is trained with a multi-task learning approach.

The encoder contains seven fully connected linear layers, each accompanied with batch normalization and \texttt{LeakyReLU} activation \citep{xu2015empirical}. The 7D input vector undergoes a sequential expansion, reaching a width of $2048$ neurons before entering the decoder stage.
To mitigate overfitting, we implement dropout \citep{JMLR:v15:srivastava14a}. Dropout later serves the purpose of isolating the epistemic uncertainty associated with the model's parameters, utilizing the Monte Carlo dropout technique \citep{gal2016dropout}. 

The encoder-decoder design serves a dual purpose: facilitating a vector-to-sequence conversion through the convolution of encoded representations, and, in a multi-task framework like ours, encouraging decoder heads to delineate features specific to the individual correlation function estimators, while the shared encoder captures features of the underlying HOD model.
An initial bottleneck linear layer adjusts the encoded representation to a width of $200$ neurons prior to entering the convolutional layers.  The model features four convolution layers with batch normalization and \texttt{LeakyReLU} activation. 
Each decoder head outputs a $40D$ vector comprising at each bin and their accompanying variances. To ensure variances are strictly positive, they are passed through a \texttt{softplus} activation in the output layer. A diagram of the full model pipeline is shown in Figure \ref{fig:pipeline}.

\subsection{Training}
\label{sec:training}

Since the decoder heads predict a distribution over the correlation function values, the model is trained with Gaussian negative log-likelihood loss \citep{MVE}. To simultaneously optimize for predicting all three correlation functions, the losses are summed. 
It is thus critical that the scale of all loss terms are roughly equal to ensure uniform learning.



We apply normalization on the inputs and each output to scale them to have zero mean and unit variance. $\xi(r)$ can exhibit strong correlations at low $r$, reaching amplitudes of $O(1000)$ or higher. $\omega(r)$ and $\eta(r)$ are however significantly noisier and have amplitudes several orders of magnitude smaller than $\xi(r)$. Amplitudes are minuscule at high $r$ for all correlation functions. Standard normalization is typical in deep learning but is especially important here due to the large variance in magnitude of the correlation amplitudes.




We train and validate our model with the AdamW optimizer \citep{loshchilov2019decoupled}, with hyperparameter tuning to optimize performance. Optimal parameters include a batch size of $128$, a step learning rate scheduler (10\% decay at 500-epoch intervals for a starting lr = 0.01), and for 1500 epochs with early stopping to discourage overfitting. We additionally employ L2-regularization via a weight decay factor of $10^{-4}$ in the optimizer. All training was conducted on one NVIDIA A100-80GB GPU.

In order to obtain epistemic uncertainties for the model predictions, we employ Monte Carlo dropout during inference. This method involves conducting several forward passes for a prediction with dropout layers left on, so that there is variation in the networks predictions. After 20 passes, we can obtain the model outputted mean and accordingly isolate the variance as the epistemic variance. A dropout rate of $0.2$ is used throughout the encoder and decoder.

\section{Results}
\label{sec:results}

\begin{figure}
    \centering
    \includegraphics[width=\textwidth]{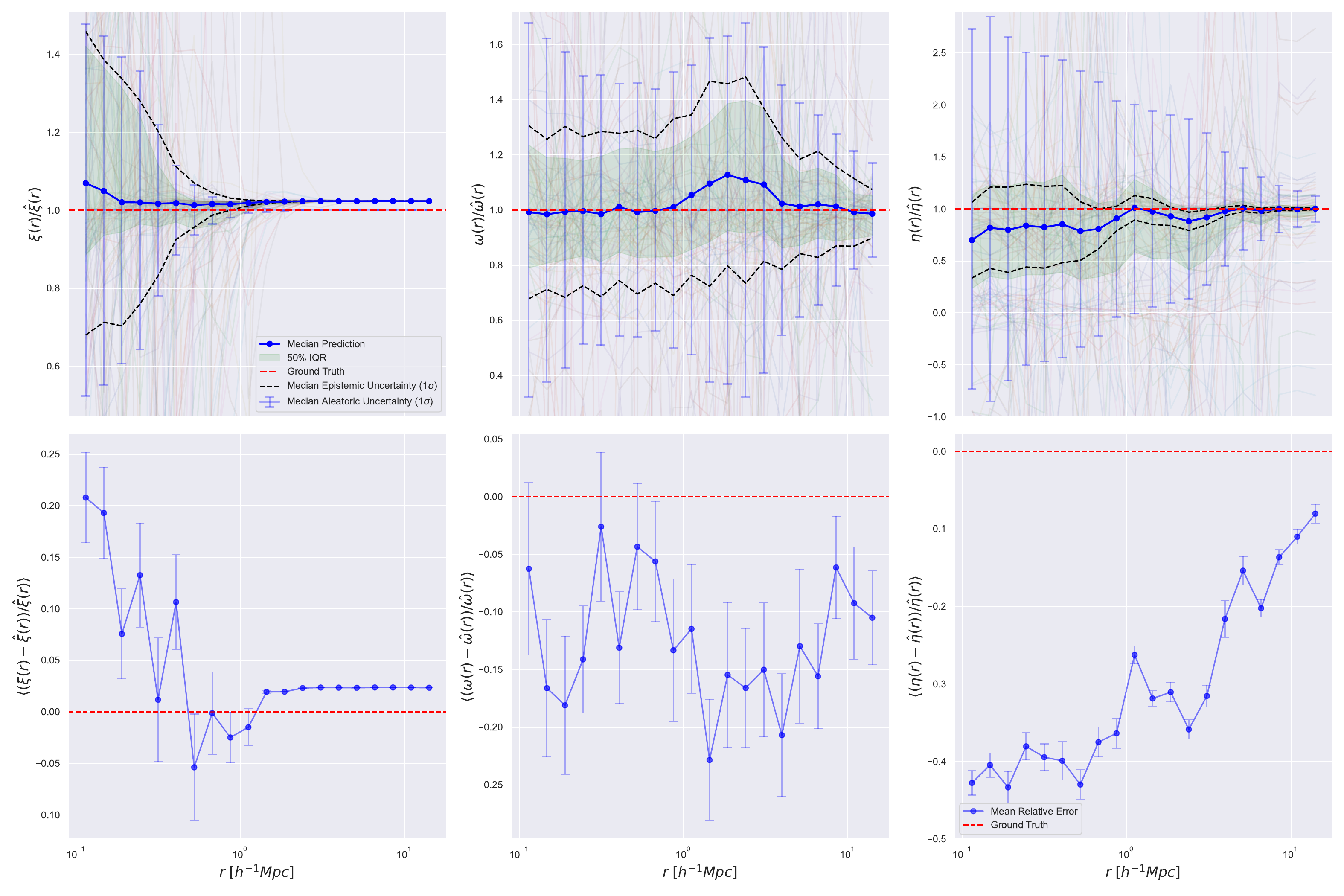}
    \caption{\textbf{Top:} Plot of median errors with aleatoric and epistemic $1\sigma$ regions of the median predictions for position-position ($\xi(r)$), position-orientation ($\omega(r)$), and orientation-orientation ($\eta(r)$) correlation functions in the scaled domain. The 50\% IQR is also shown in green. The results for a random sample of 100 test-set predictions are shown in the background, showcasing the variability of performance. \textbf{Bottom:} Mean fractional errors with aleatoric uncertainties for $\xi(r)$, $\omega(r)$, and $\eta(r)$. Ratios are computed with respect to the running mean using a window size of 3 for $\omega(r)$ and $\eta(r)$ due to their increased noise.}
    \label{fig:results}
\end{figure}

\paragraph{Accuracy} We evaluate the model performance on the 20\% test set. It can be seen in Figure \ref{fig:results} that the median errors of $\xi(r)$ and $\omega(r)$ (blue line) are well-behaved and near the zero error benchmark (red-dashed line). Further, the 50\% interquartile range (IQR) shows shows a uniform variability around the median error for $\omega(r)$ and $\eta(r)$ and includes an error of zero. For $\xi(r)$, the 50\% IQR is skewed high at low $r$, indicating a bias. The median error of $\eta(r)$ is $\sim 25\%$ at low $r$ and $\leq 10\%$ in the 2-halo regime. Outlier predictions are also shown in Figure \ref{fig:results}, showcasing the variability in performance at low $r$ for $\xi(r)$ and across all scales for $\eta(r)$ and $\omega(r)$. 

In general, model accuracy and confidence increases with higher $r$, when most correlations are small. For $\xi(r)$, at low $r$ model predictions are biased high and there is high aleatoric uncertainty. The fractional error is generally $\leq 15\%$, and the model achieves $\sim 2.5\%$ accuracy at high $r$. The statistics $\omega(r)$ and $\eta(r)$ exhibit significantly higher noise levels, resulting in larger aleatoric uncertainties. Preliminary testing reveals that the average uncertainties closely align with those extracted directly from the simulation. For $\omega(r)$, the fractional errors are generally $\leq 20\%$, and the model's predictions tend to be biased low. Moreover, for $\omega(r)$ and $\eta(r)$, the ground truth falls within the 50\% IQR of the median prediction across all bins.

We find strong correlations between model predictions and the respective ground truths as characterized by the Pearson correlation coefficient (PCC), which quantifies agreement between the overall shapes of the ground truth and model predictions. We find that the mean $PCC(\xi) = 0.98$, $PCC(\omega) = 0.88$, and $PCC(\eta) = 0.65$. The fractional errors of the model do not indicate a particularly strong performance in the case of $\omega(r)$ and $\eta(r)$; these metrics have potential to be inflated due to the small amplitudes and large stochasticity of these correlations, even when using methods such as computing the running mean to illustrate the errors as shown in Figure \ref{fig:results}. Example model predictions for a variety of scenarios is further shown in Appendix \ref{sec:appendixA}. We emphasize in this case that it was the intention for the model to capture this underlying signal of the correlations and not overfit to noise, which is supported by the PCC values. 
\paragraph{Calibration} We  find that $45.18\%$ of $1\sigma$ epistemic uncertainties are smaller than aleatoric uncertainties for $\xi(r)$, indicating that a larger training sample would be of benefit. $98.22\%$ and $99.72\%$ are smaller than the aleatoric uncertainty at $1\sigma$ for $\omega(r)$ and $\eta(r)$. For $\xi(r)$, we find that $40.32\%$ and $47.23\%$ of predictions fall within $1\sigma$ and $2\sigma$ of their ground truth values, respectively. This low percentage is largely due to the highly confident, but systemically biased predictions for correlations at large $r$ as seen in the fractional error of $\sim 2.5\%$ in Figure \ref{fig:results}.  As seen in Figure \ref{fig:results}, the high stochasticity in $\omega(r)$ and $\eta(r)$ has resulted in large model-predicted aleatoric uncertainties. For $\omega(r)$, the corresponding percentages are $85.80\%$ within $1\sigma$ and $98.82\%$ within $2\sigma$. For $\eta(r)$, $77.43\%$ of predictions fall within $1\sigma$ and $97.35\%$ within $2\sigma$. These confidence intervals indicate that the aleatoric uncertainty is slightly inflated; nevertheless, for a more robust conclusion, we plan to to conduct a thorough validation by directly extracting the uncertainty from the simulation data itself by running multiple realizations for each set of parameters.

\paragraph{Limitations} 
We are largely limited by the sparse signal present in inherently noisy correlations such as $\omega(r)$ and $\eta(r)$. The model can capture the underlying signal for these two correlations but accordingly predicts a large uncertainty, particularly for low $r$. This was indeed the desired outcome -- that the model would learn to predict the ``cosmic mean'' in the presence of significant noise. At high $r$, when correlations are small, the median predictions are largely correct and confident, aside from the $\sim2.5\%$ systemic bias in $\xi(r)$. The PCC values indicate that despite considerable error in correlation amplitudes, the model successfully captures the long-range behavior of the correlations well.

\paragraph{Previous Considerations} We previously tested other data normalization schemes such as min-max normalization and custom scaling, wherein we defined $\xi' = \xi / \xi_{DM}$ and scaled $\omega(r)$ and $\eta(r)$ as

\begin{equation}
    \omega'_i = \frac{\mu_{\xi'}}{\mu_{\omega}} \cdot \omega_i  \quad \quad \quad
    \eta'_i = \frac{\mu_{\xi'}}{\mu_{\eta}} \cdot \eta_i
\end{equation}

where $\mu$ denotes the mean. This normalization was extensively studied as it includes some information of the underlying dark matter distribution, which we believed would assist the learning. However, we found that it disadvantaged $\omega(r)$ in the 1-halo regime and did not work as well as standard normalization. Interestingly, the bias in $\xi(r)$ predictions at high $r$ was absent with this normalization. Lastly, we studied other architectures, including fully-connected and U-Net based architectures, as well as variations to the current encoder-decoder design and found that the inclusion of 1D convolution aided performance. Single-task learning on individual correlations was also studied, in which we found that $\omega(r)$ and $\eta(r)$ performance was slightly improved but more susceptible to overfitting. Thus, the information shared among correlations in the encoder stage yields benefits over single-task learning. We additionally conducted experiments in only predicting point-estimates, and found that the inclusion of mean-variance estimation was essential to understanding the degree of stochasticity in $\omega(r)$ and $\eta(r)$ and properly quantifying their performance.

\section{Discussion}

We have presented a model that efficiently predicts galaxy IA correlations in terms of HOD simulation parameters without costly simulation. The model can perform inference on a batch of $32,768$ input parameters in $1.02$ seconds on one NVIDIA A100-80GB GPU, while the simulation when run in parallel on 150 CPU cores for the same parameters takes $\mathcal{O}(3 \; \text{hours})$. The model simultaneously predicts point estimates and uncertainties for three correlations spanning $20$ bins whose values span several orders of magnitude and are also significantly noisy, making it a notable data and model engineering task to isolate the relevant signal and not overfit to the noise when signal is sparse.
The model effectively avoids overfitting to this noise and demonstrates conservative aleatoric uncertainties, which will be validated in future work. Furthermore, it accurately captures the underlying signal of correlations as shown by the PCC values.  Point-predictions are most accurate for $\xi(r)$ ($\leq 15\%$, generally), with large fractional errors for $\omega(r)$ and $\eta(r)$ which can be inflated due to the large stochasticity of these correlations.



The epistemic uncertainty of the model is typically lower than the aleatoric uncertainty and is well-calibrated as noted by the confidence interval statistics shown in \cref{sec:results}. The epistemic uncertainty is something we aim to decrease in future work. This can be mitigated by conducting multiple realizations for parameters to curate a larger training sample, or alternatively by decreasing the number of parameters in the encoder stage of the model \citep{Semenova_2022}, though typically at the expense of model expressivity. In doing so, we can improve the quality of model predictions, further calibrate its aleatoric uncertainty predictions, and simultaneously provide a benchmark for which to validate them with.

\subsection{Future Work}

Several improvements can be made at both a data engineering and architectural level. Of particular importance is understanding the true degree of stochasticity in $\omega(r)$ and $\eta(r)$ and addressing the systemic bias for $\xi(r)$ in the 2-halo regime.
We also plan to study different parameter configurations for generating data and further investigate combinations which resulted in unphysical universes. We plan to perform parameter inference on $\mu_{sat}$ and $\mu_{cen}$, as well as other galaxy occupation parameters, with the model utilizing Markov Chain Monte Carlo (MCMC) once the pipeline is refined. The simulation itself is very representative of real data; the eventual goal is to 
validate the model with the IllustrisTNG suite of simulations, similar to what was done in \citet{vanalfen_2023}, as well as real data. We additionally plan to consider more complex HOD model dependence, such as distance-dependence alignment. The eventual goal is to generalize well beyond the cosmological parameters that are implicit in the models training data, such as those that determine the underlying cosmological simulation, with hopes of creating a unifying and efficient emulator which will accelerate the study of IA correlations and their effects on cosmological measurements.

\section{Reproducibility Statement}

The entire procedure of this work, from data generation to modeling and evaluation, are able to reproduced. The data generation procedure can be reproduced using \texttt{halotools} (\url{https://halotools.readthedocs.io}) with the appropriate parameters and literature outlined in \cref{sec:sim}. A description of the encoder-decoder architecture that was written in \texttt{PyTorch} is summarized in \cref{sec:arch} and \cref{sec:training}, with a detailed description of the architecture further provided in Appendix \ref{sec:AppendixB}. Our code will be made open source upon journal publication.

\section{Acknowledgements}

We thank the anonymous referees for their useful comments. S.P. acknowledges support from the National Science Foundation under Cooperative Agreement PHY-2019786 (The NSF AI Institute for Artificial Intelligence and Fundamental Interactions, \url{https://iaifi.org}). Y.Y. acknowledges support from Khoury Apprenticeship program. J.B. and N.V.A. are supported in this work by NSF award AST-2206563 and the Roman Research and Support Participation program under NASA grant 80NSSC24K0088. R.W. is supported by NSF award DMS-2134178. Data generation was conducted on the Discovery cluster, supported by Northeastern University’s Research Computing team. The machine learning computations were run on the FASRC cluster supported by the FAS Division of Science Research Computing Group at Harvard University.




\newpage
\bibliography{paper}

\newpage
\begin{appendices}
\section{Model Predictions}
\label{sec:appendixA}


\begin{figure}[htbp!]
    \centering
    \includegraphics[width=.89\textwidth]{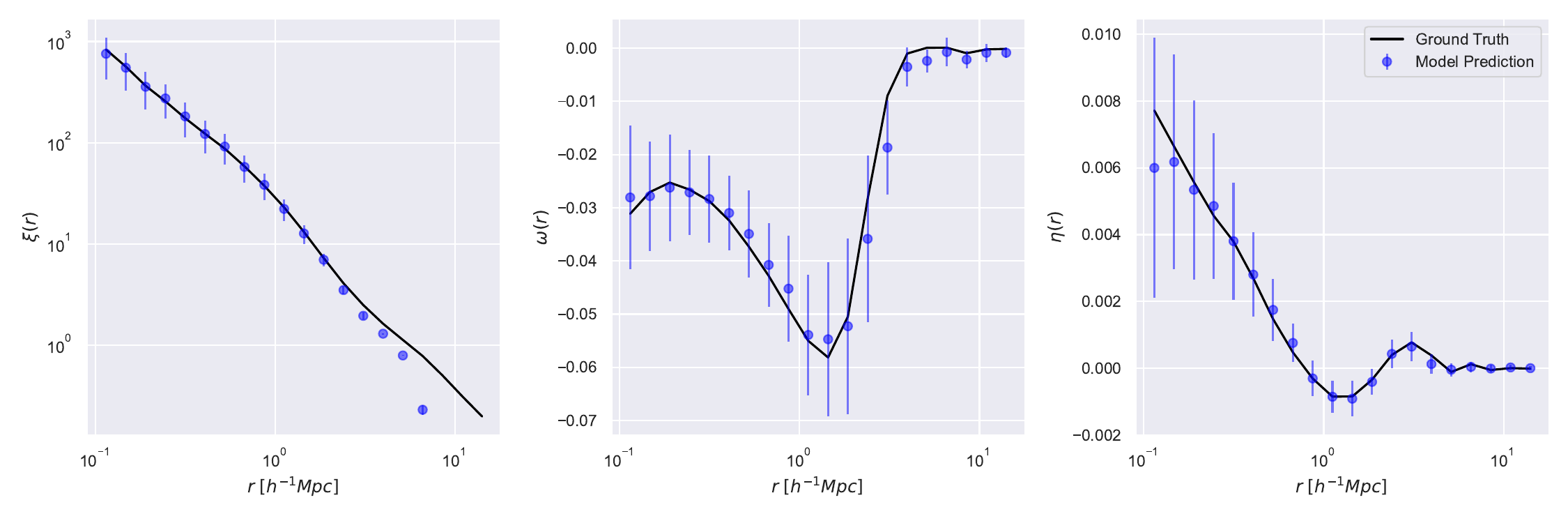}
    \caption{Good performance examples on test set in rescaled domain. On the left panel one can see the small bias in $\xi(r)$ predictions at high $r$. }
    \label{fig:good_exs}
\end{figure}

\begin{figure}[htbp!]
    \centering
    \includegraphics[width=.89\textwidth]{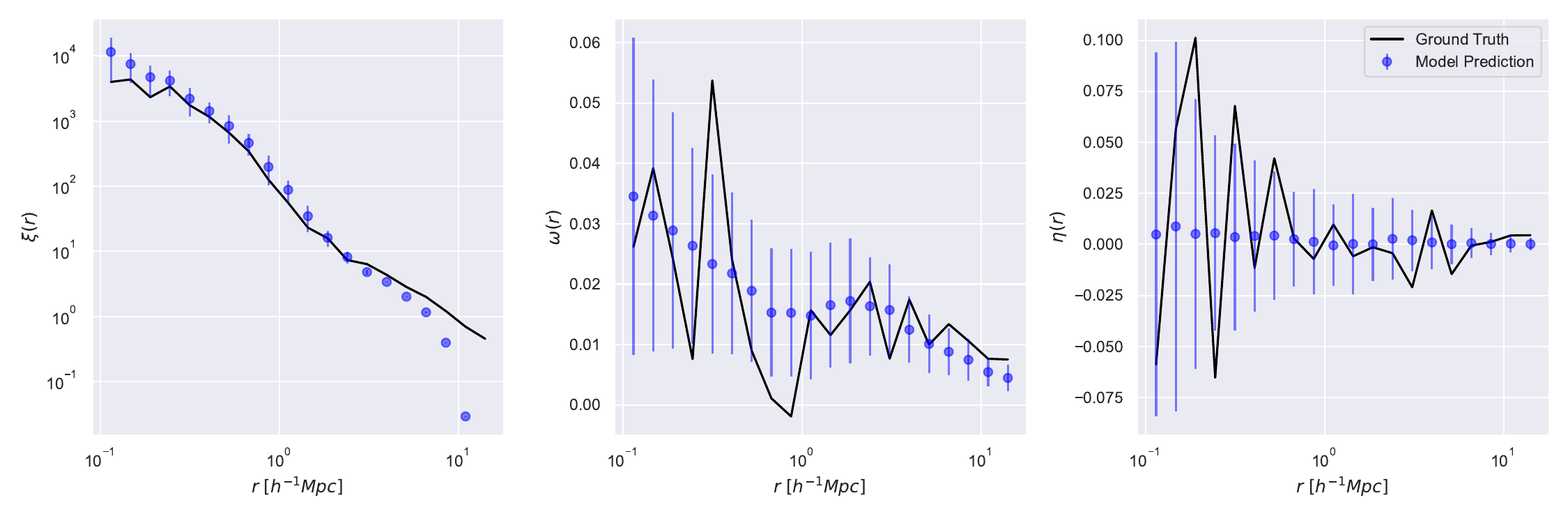}
    \caption{Noisy data examples on test set in rescaled domain. $\xi(r)$ exhibits a small bump at low $r$ which is likely a noisy artifact. $\omega(r)$ and $\eta(r)$ are very noisy, but the model is able to capture the underlying signal as well as overall shape.}
    \label{fig:noisy_exs}
\end{figure}

\begin{figure}[htbp!]
    \centering
    \includegraphics[width=.89\textwidth]{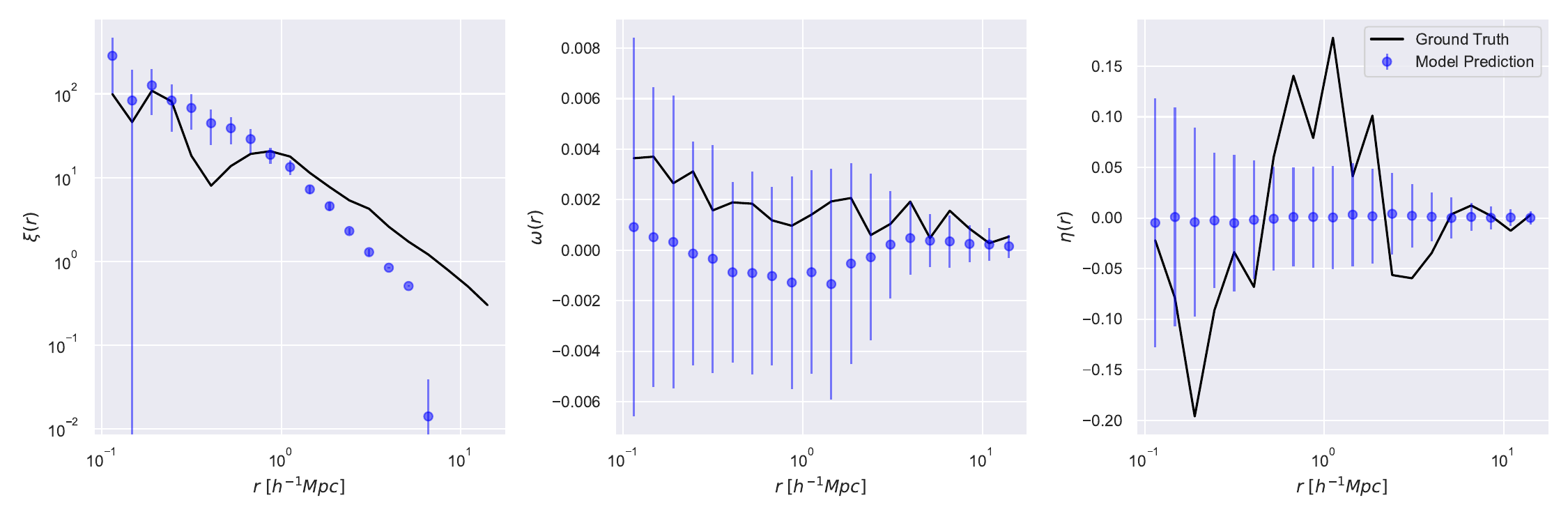}
    \caption{Poor performance examples on test set in rescaled domain.}
    \label{fig:bad_exs}
\end{figure}



\section{Model Architecture}
\label{sec:AppendixB}

\begin{table}[h!]
    \centering
    \begin{tabular}{|ccccc|}
        \hline
        \textbf{Layers} & \textbf{Properties} & \textbf{Stride} & \textbf{Padding} & \textbf{Output Shape} \\
        \hline
        \multicolumn{5}{|c|}{Encoder} \\
        \hline
        Input: 7 & & & &\\
        \hline
        Linear & Width: 128 & & &\\
        (w/ BatchNorm1D) & Dropout: 0.2 & - & - & (, 128) \\
        & Activation: LeakyReLU & & &\\
        \hline
        Linear & Width: 256 & & &\\
        (w/ BatchNorm1D) & Dropout: 0.2 & - & - & (, 256) \\
        & Activation: LeakyReLU & & &\\
        \hline
        Linear & Width: 512 & & &\\
        (w/ BatchNorm1D) & Dropout: 0.2 & - & - & (, 512) \\
        & Activation: LeakyReLU & & &\\
        \hline
        Linear & Width: 1024 & & &\\
        (w/ BatchNorm1D) & Dropout: 0.2 & - & - & (, 1024) \\
        & Activation: LeakyReLU & & &\\
        \hline
        Linear & Width: 2048 & & &\\
        (w/ BatchNorm1D) & Dropout: 0.2 & - & - & (, 2048) \\
        & Activation: LeakyReLU & & &\\
        \hline
        \multicolumn{5}{|c|}{Decoder \; (x3)} \\
        \hline
        Linear (Bottleneck) & Width: 200 & & & (, 200)\\
        \hline
        Conv1D & Filters: 20 & & &\\
        (w/ BatchNorm1D)& Kernel: 3x3 & 1 & 1 & (20, 10)\\
        & Activation: LeakyReLU & & &\\
        & Dropout: 0.2 & & & \\
        \hline
        Conv1D & Filters: 40 & & &\\
        (w/ BatchNorm1D)& Kernel: 3x3 & 1 & 1 & (40, 10)\\
        & Activation: LeakyReLU & & &\\
        & Dropout: 0.2 & & & \\
        \hline
        Conv1D & Filters: 80 & & &\\
        (w/ BatchNorm1D)& Kernel: 5x5 & 1 & 2 & (80, 10)\\
        & Activation: LeakyReLU & & &\\
        & Dropout: 0.2 & & & \\
        \hline
        Conv1D & Filters: 20 & & &\\
        (w/ BatchNorm1D)& Kernel: 3x3 & 5 & 1 & (20, 2)\\
        & Activation: LeakyReLU & & &\\
        \hline
        Output: (, 3, 40) & & & &\\
        \hline
    \end{tabular}
    \label{table:cnn_table}
\end{table}

\end{appendices}

\end{document}